\documentclass[12pt]{article}
\date{}
\title{Neutrino Bremsstrahlung Process in highly degenerate magnetized electron gas}
\author{Indranath Bhattacharyya\\
Department of Applied Mathematics\\
University of Calcutta, Kolkata-700 009, INDIA\\E-mail :
$i_{-}bhattacharyya@hotmail.com$\\} \vskip .1in
\begin{document}
\maketitle \baselineskip .3in \noindent
\begin{abstract}
In this article the neutrino bremsstrahlung process is considered
in presence of strong magnetic field, though the calculations for
this process in absence of magnetic field are also carried out
simultaneously. The electrons involved in this process are
supposed to be highly degenerate and relativistic. The scattering
cross sections and energy loss rates for both cases, in presence
and absence of magnetic field, are calculated in the
extreme-relativistic limit. Two results are compared in the range
of temperature $5.9\times 10^{9}$ K $< T\leq 10^{11}$ K and
magnetic field $10^{14} - 10^{16}$ G at a fixed density $\sim
10^{15}$ $gm/cc$, a typical environment during the cooling of
magnetized neutron star. The interpretation of our result is
briefly discussed and the importance of this process during the
stellar evolution is speculated.
\end{abstract}
\section{Introduction:}
\indent The neutrino emission process plays an important role in
the late stages of the stellar evolution. It is known that the
radiation of neutrino could be dominated over the ordinary
electromagnetic radiation in case of highly dense and hot stellar
structures, such as, white dwarves or neutron stars. Unlike the
other mechanisms the neutrinos are produced directly from their
point of origin and do not require the transport of energy to the
surface of the stellar object before getting radiated. As a
consequence the energy outflow is given directly by the rate at
which the neutrinos are produced. Photo-neutrino process
$(e^{-}+\gamma\longrightarrow \nu+\overline{\nu})$,
Pair-annihilation process $(e^{-}+e^{+}\longrightarrow
\nu+\overline{\nu})$ and Plasma neutrino process
$(\Gamma\longrightarrow \nu+\overline{\nu})$ are the sole
mechanisms that carry away the energy from star during its
evolution period, although there are few more processes which
might play more important role under some special environments.
Chiu and his collaborators \cite {Chiu1960, Chiu1961, Chiu}
calculated some important neutrino emission processes and pointed
out their important role in astrophysics. In 1972 Dicus
\cite{Dicus1972} reconsidered a few such processes in the
framework of electro-weak interaction theory and calculated the
energy-loss rate at the various stages of the stellar evolution.
According to the Standard Model the neutrino has the minimal
properties such as zero mass, zero charge etc. Introduction of
neutrino mass compatible with the experimental data is a bare
minimum extension of the Standard Model. In the frame work of
Standard Model with this little extension some calculations
\cite{Dicus2000, Dodelson} related to the neutrino process have
already been carried out. Itoh et al. \cite{Itoh} considered some
neutrino emission processes to calculate them numerically.\\
\indent Pontecorvo \cite{Pontecorvo} gave an idea that the
neutrino may be emitted by the interaction of electron with the
nucleus, called bremsstrahlung process. Gandel'man and Pinaev
\cite{Gandel'man} carried out the detailed calculations for this
process in the non-relativistic electron gas. After that Festa and
Ruderman \cite{Festa} extended this calculations for relativistic
limit with considering the screening effect that becomes important
in the high density. Dicus et al. \cite{Dicus} considered the
process according to the Standard Model and compare the result
with different screening effect. Saha \cite{Saha} calculated the
bremsstrahlung process according to photon-neutrino weak coupling
theory which is very much satisfactory in explaining the
neutrino-synchrotron process \cite{Raychaudhuri1970}. In this
article we have considered the bremsstrahlung process in presence
of strong magnetic field and this is the first attempt to do so.
It has been shown by Festa and Ruderman \cite{Festa} and then by
Dicus et al. \cite{Dicus} that the bremsstrahlung process has
maximum effect in the relativistic degenerate region; thus in the
stellar object such as newly born born neutron star the process is
supposed to be very much effective. Here we are to verify whether
the presence of strong magnetic field, which may be generated in
the rotating neutron stars, will have any effect on the
bremsstrahlung process. We have calculated the scattering cross
section and then obtained the energy-loss rate for the
bremsstrahlung both in presence and absence of strong magnetic
field. We are going to study the influence of high magnetic field
on the bremsstrahlung process and the region where its presence
may take a crucial role in the neutrino emission. The ordinary
neutrino bremsstrahlung process is very much significant for the
neutrino energy generation process in the relativistic highly
degenerate region; therefore a comparative study is required for
this process in presence and also in absence of strong magnetic
field.
\section{Calculation of scattering cross section:}
In the bremsstrahlung process the electron interacts with the
nucleus having the coulomb potential
$$\Phi(\overrightarrow{r})=\frac{Ze}{\mid\overrightarrow{r}\mid}
e^{-\mid\overrightarrow{r}\mid /\lambda_{d}}\eqno{(2.1)}$$ i.e. a
single charge with an exponential screening cloud (Yukawa like
charge distribution), where $\lambda_{d}$ is the Debye screening
length given by
$$\frac{1}{\lambda_{d}^{2}}=\frac{4e^{2}}{\pi}E_{F}P_{F}\eqno{(2.2)}$$
Here $P_{F}$ and $E_{F}$ represent the Fermi momentum and energy
respectively. We consider the potential, given by (2.1), because
the screening effect is important in the high density region.
There will be four different Feynman Diagrams shown in the
Figure-1 (Z-exchange diagram) and Figure-2 (W-exchange diagram).
In our calculations the presence of magnetic field plays a crucial
role, so it is to be handled with care. Without any loss of
generality we can take the direction of magnetic field along
$z$-axis. In presence of magnetic field the energy momentum
relation of the electron becomes
$$(p_{n}^{0})^{2}=m_{e}^{2}+ p_{z}^{2}+2n\frac{H}{H_{c}}m_{e}^{2}\eqno{(2.3)}$$
where $H_{c}=4\cdot 414$ G stands for critical magnetic field.
Here $n$ represents the Landau level for the electron in the
magnetic field. [The value of $s$ is taken as $\pm 1$ as per the
spin of the electron is directed towards or opposite to the
direction of magnetic field (along z-axis) respectively.]
\\\indent The component of electron momentum along the direction of
magnetic field remains unaffected. It is clear that the effect of
magnetic field on the electron quantizes its energy to the
direction perpendicular to $H$ and thus transverse components
would get replaced by
$p_{x}^{2}+p_{y}^{2}\longrightarrow2nm_{e}^{2}\frac{H}{H_{c}}$,
whereas the longitudinal component $p_{z}$ would be directed along
the magnetic field. The Feynman diagrams for the bremsstrahlung
process are same for both in presence and absence of magnetic
field; only we have to keep in our mind that four momenta of the
electronic lines, present in the diagrams, should be modified.
First we shall calculate the ordinary bremsstrahlung process i.e.
the process in absence of magnetic field. The matrix element can
be constructed as
$$M_{fi}=-ie\frac{G_{F}}{\sqrt{2}}A_{0}(\overrightarrow{k})J_{\mu}\mathcal{M}^{\mu}\eqno{(2.4)}$$
where,
$$\mathcal{M}^{\mu}=\overline{u}(p')[\gamma^{\mu}(C_{V}-C_{A}\gamma_{5})
\frac{(p^{'\tau}\gamma_{\tau}+q^{\tau}\gamma_{\tau}+m_{e})}
{(p'+q)^{2}-m_{e}^{2}+i\epsilon}\gamma^{0}+\gamma^{0}
\frac{(p^{\tau}\gamma_{\tau}-q^{\tau}\gamma_{\tau}+m_{e})}
{(p-q)^{2}-m_{e}^{2}+i\epsilon}\gamma^{\mu}(C_{V}-C_{A}\gamma_{5})]u(p)\eqno{(2.5)}$$
$$J_{\mu}=\overline{u}_{\nu}(q_{1})\gamma_{\mu}(1-\gamma_{5})v_{\nu}(q_{2})\eqno{(2.6)}$$
$$A_{0}(\overrightarrow{k})=\int\Phi(\overrightarrow{r})
e^{-(\overrightarrow{k}.\overrightarrow{r})}d^{3}r=-\frac{4\pi
Ze}{\mid k^{2}+q_{sc}^{2}\mid}\eqno{(2.7)}$$ and $$q=q_{1}+q_{2}$$
The energy momentum conservation leads to
$$k+p=p'+q$$ where $k$ is purely space like as in the case of photo-coulomb
neutrino process \cite{Rosenberg, Bhattacharyya1}. The term
$q_{sc}$ present in the equation (2.7) arises due to the screening
effect and can be expressed as
$$q_{sc}=\frac{1}{\lambda_{d}}\eqno{(2.8)}$$
Using some simplifications we can write the term
$J_{\mu}\mathcal{M}^{\mu}$ as follows:
$$J_{\mu}\mathcal{M}^{\mu}=[\frac{(p'J)}{q^{2}/2+(p'q)}+\frac{(pJ)}{q^{2}/2-(pq)}]
\overline{u}(p')\gamma^{0}(C_{V}-C_{A}\gamma_{5})u(p)\eqno{(2.9)}$$
We can put this expression to the equation (2.4) to get the
expression for the scattering matrix. Now the squared sum of the
scattering matrix over the final spins is to be integrated over
the final momenta. The squared sum of the expression
$[\overline{u}(p')\gamma^{0}(C_{V}-C_{A}\gamma_{5})u(p)]$ gives
$$\sum\mid\overline{u}(p')\gamma^{0}(C_{V}-C_{A}\gamma_{5})u(p)\mid^{2}=(C_{V}^{2}-C_{A}^{2})
+(C_{V}^{2}+C_{A}^{2})\frac{(p'_{0}p_{0}-\overrightarrow{p}'.\overrightarrow{p})}{m_{e}^2}\eqno{(2.10)}$$
Note that in the equation (2.10) no neutrino momentum is present,
so this part will not be taken into account during the integration
over the final momenta of neutrinos. Let us now evaluate the
squared spin sum of the expression
$\mid\frac{(p'J)}{q^{2}/2+(p'q)}+\frac{(pJ)}{q^{2}/2-(pq)}\mid$
and then integrating over final momenta of the neutrinos we can
obtain [See Appendix-A]
$$\int\sum\mid\frac{(p'J)}{q^{2}/2+(p'q)}+\frac{(pJ)}{q^{2}/2-(pq)}\mid^{2}
\frac{d^{3}q_{1}d^{3}q_{2}}{(2\pi)^{3}2q^{0}_{1}(2\pi)^{3}2q^{0}_{2}}(2\pi)
\delta(q^{0}-q^{0}_{1}-q^{0}_{2})$$
$$\approx\frac{1}{18(2\pi)^{3}m_{\nu}^{2}}(p^{0}-p^{'0})^{3}
[\frac{\mid\overrightarrow{p}-\overrightarrow{p}'\mid}{p^{0}+p^{'0}}]^{2}\eqno{(2.11)}$$
Up to this step the calculations for the neutrino bremsstrahlung
process would be same in both situations i.e. in presence as well
as absence of magnetic field. It is assumed that in presence of
magnetic field quantized transverse components of the momentum do
not participate directly during the interaction between nucleus
and electron. Only the z-component of the electron momentum takes
part in this process. Thus in this case we can obtain an
expression almost similar to the equation (2.11) with replacing
$\mid\overrightarrow{p}'\mid$ and $\mid\overrightarrow{p}\mid$ by
$p'_{z}$ and $p_{z}$ respectively. Now to integrate the squared
sum of the matrix element over all final momenta we shall utilize
the result obtained in the equation (2.10) and (2.11), but we have
to take care when the magnetic field is present. In this case the
phase space factor $d^{3}p'$ takes the form [See Appendix-B]
$$\int d^{3}p'=\pi \frac{H}{H_{c}}m_{e}^{2}\int dp_{z}'\eqno{(2.12)}$$
whereas in the ordinary bremsstrahlung process the integration
over the final momentum of electron is done in the usual manner.
In the center of mass frame and assuming the electron momentum is
much high relative to its rest mass we can carry out the
calculations. In this extreme relativistic limit we can evaluate
the integral over the final momentum of electron and obtain the
following expression.
$$\int\sum\mid M\mid^{2}\frac{d^{3}q_{1}d^{3}q_{2}d^{3}p'}
{(2\pi)^{3}2q^{0}_{1}(2\pi)^{3}2q^{0}_{2}(2\pi)^{3}2p^{'0}}(2\pi)\delta(q^{0}-q^{0}_{1}-q^{0}_{2})
=\frac{8G_{F}^{2}\alpha^{2}Z^{2}}{9(2\pi)^{3}(1+r^{2})^{2}}(C_{V}^{2}+C_{A}^{2})\frac{p_{0}^{3}}
{m_{e}^{2}m_{\nu}^{2}}\eqno{(2.13)}$$ This expression is obtained
for the ordinary bremsstrahlung process when there will be no
magnetic field. In presence of magnetic field this expression
becomes
$$\frac{2G_{F}^{2}\alpha^{2}Z^{2}}{9(2\pi)^{3}}(C_{V}^{2}+C_{A}^{2})\frac{p_{0}}
{m_{\nu}^{2}}(\frac{H}{H_{c}})\eqno{(2.14)}$$ Note that the term
$r$ arises due to the weak screening effect. It is given by
$$r\approx \frac{q_{sc}}{p^{0}}$$
 The expression for $C_{V}$ and $C_{A}$ for the electron type of
neutrino emission will differ from those in case of muon and tau
neutrino emission, since W-boson exchange diagrams are present
only when the electron neutrino anti-neutrino pair is emitted.
Inserting the above terms into the expression of the scattering
cross section for both of those cases and returning to the C.G.S.
unit we can finally obtain
$$\sigma\approx1\cdot76\times10^{-50}(\frac{E}{m_{e}c^{2}})^{2}\frac{1}{(1+r^{2})^{2}}
\hspace{0.5cm}cm^{2}\eqno{(2.15)}$$ in absence of magnetic field,
whereas
$$\sigma_{mag}\approx4\cdot41\times10^{-51}(\frac{H}{H_{c}})
\hspace{0.5cm}cm^{2}\eqno{(2.16)}$$ in presence of magnetic field.
\\ It is worth noting that all three type of neutrinos are taken into account in our
calculations. Our result (equation 2.16) shows that the scattering
cross section for the bremsstrahlung process in presence of
magnetic field will not depend on the energy of the incoming
electron, but on the intensity of the magnetic field present in
the surroundings.
\section{Calculation of energy loss rate:}
In the extreme relativistic case the energy loss rate in erg per
nucleus per second for the neutrino bremsstrahlung process is
calculated by the formula
$$\mathcal{E}_{\nu}^{Z}=\frac{2}{(2\pi)^{3}\hbar^{3}}
\int\frac{d^{3}p}{[e^{\frac{E-E_{F}}{\kappa T}}+1]}c\sigma E
e^{\frac{E-E_{F}}{\kappa T}}\eqno{(3.1)}$$ where $E_{F}$ stands
for Fermi energy of the electron. We are considering the case in
which the electrons are highly degenerate. It is well known that
in the degenerate region the energy of the electron remains below
the Fermi energy level. To obtain the energy-loss rate in erg per
gram per second $\mathcal{E}_{\nu}^{Z}$ is divided by $Am_{p}$ and
it is obtained as
$$\mathcal{E_{\nu}}=\frac{Z^{2}}{A}\times5\cdot 26\times10^{-3}\times
\frac{x_{F}^{6}e^{1-x_{F}}} {(1+r^{2})^{2}}T_{10}^{6}\hspace{0.5
cm}erg/gm-sec\eqno{(3.2)}$$ where, $T_{10}=T\times 10^{-10}$\\The
$x_{F}$ represents the ratio of the Fermi temperature to the
maximum temperature of the degenerate electron gas. The degeneracy
will be attained only when the following condition will be
satisfied \cite{Chandrasekhar}.
$$x_{F}^{2}>2\pi^{2}\eqno{(3.3)}$$
It can be obtained $x_{F}\approx 6$, considering the fact that
temperature and density of the electron gas present in the core of
a newly born neutron star would be approximately $10^{12}$ K and
$10^{15}$ $gm/cc$ respectively. The term $r$ arises due to the
weak screening, related to $x_{F}$ by $r\approx 0\cdot 096 \times
x_{F}$. Thus we can calculate the term $r$, present in the
equation (3.2). Finally the expression for the energy-loss rate
becomes
$$\mathcal{E_{\nu}}=\frac{Z^{2}}{A}\times0\cdot93\times10^{12}\times
T_{10}^{6} \hspace{0.5 cm}erg/gm-sec\eqno{(3.4)}$$ In the same
manner we can obtain the energy loss rate in presence of magnetic
field. In that case the the phase space factor will be replaced
according to the rule defined in (B6). In the same manner
energy-loss rate can be calculated as
$$\mathcal{E_{\nu}}^{mag}=\frac{Z^{2}}{A}\times0\cdot51\times10^{6}
\times H_{13}^{2}\times
T_{10}^{2}\hspace{0.5cm}erg/gm-sec\eqno{(3.5)}$$ where,
$H_{13}=H\times 10^{-13}$ \\We have computed (Table-1) the
logarithmic value of the energy loss rate in the temperature range
$0\cdot 8\times 10^{10} - 10^{11}$ K and the magnetic field
$10^{14}- 10^{16}$ G at a fixed density $\rho=10^{15}$ $gm/cc$.
\section{discussion:}
The neutrino bremsstrahlung process is an important energy
generation mechanism during the stellar evolution and very much
effective in the highly degenerate region, for examples, in the
cores of low mass red giants, white dwarves etc. In addition to
this degenerate nature if the electron gas is highly relativistic
the energy-loss rate through the bremsstrahlung process would be
significantly high. It has already been calculated that the
neutrino luminosity in the crust of neutron star is high enough
\cite{Maxwell}, but it is yet to be verified what would be the
effect of neutrino emission by the bremsstrahlung process in the
core region, particularly when the core is strongly magnetized.
The discovery of radio pulsars showed that the collapse of normal
stars results not only in supernova explosions, but may also
generate strong magnetic field. In some stellar objects like
neutron stars and magnetars the magnetic field may reach to
$10^{16}$ G and influences the neutrino emission process. It is
known that the neutron star is born as a result of type II
supernova explosion. In the newly born neutron star the core
temperature becomes $10^{12}$ K which drops down to $10^{11}$ K
within few seconds of its birth and then slowly cools down until
the temperature reaches to $2\times10^{8}$ K, after which the
electromagnetic emission dominates over the neutrino emission
\cite{Raffelt}. It is worth noting that the electron gas, still
left in the stellar core, is highly degenerate and relativistic.
\\\indent We are going to verify the role of the magnetic field
on the neutrino bremsstrahlung process. In the neutrino
synchrotron radiation \cite{Landstreet, Raychaudhuri1970, Canuto,
Bhattacharyya2} neutrino anti-neutrino pair emission takes place
since the electron changes its Landau levels, but in the
bremsstrahlung the Landau levels are assumed to be unchanged
throughout the process. The process occurs through the change of
magnitude of the component of electron linear momentum directed
along the magnetic field. It is found from Table-1 that in the
temperature range $10^{10}$ K $\leq T\leq 10^{11}$ K and at the
density $10^{15}$ $gm/cc$ the energy loss rate for the ordinary
bremsstrahlung process is greater than that in presence of strong
magnetic field ($10^{14} - 10^{16}$ G). We can interpret that in
the early stage of neutron star cooling, when the temperature
remains above $10^{10}$ K, the effect of the bremsstrahlung
process get lowered due to the presence of magnetic field,
although the energy loss rate is still very high. If the strength
of the magnetic field goes below the critical value, the process
would become free from the influence of magnetic field, and a
greater amount of neutrino energy is produced. It is evident from
our work that though, in general, the magnetic field makes the
bremsstrahlung process a bit less effective, but there exists a
particular region ($5\cdot 9\times 10^{9}$ K $<T< 10^{10}$ K,
$H\sim 10^{16}$ G and $\rho\sim 10^{15}$ $gm/cc$) during neutron
star cooling, where the bremsstrahlung process contributes a
greater amount of neutrino energy loss by the influence of
magnetic field compared to the situation when there would be no
magnetic field at all. Therefore, it can be predicted that if the
temperature falls below $10^{10}$ K, the process would give
maximum effect due to the presence of super strong magnetic field
having intensity $10^{16}$ G. Our study reveals that the neutrino
bremsstrahlung process is an important energy generation mechanism
in the late stages of the stellar evolution, even in presence of
magnetic field.
\section{Acknowledgement:} I am very much thankful to Prof.
{\bf Probhas Raychaudhuri} of The Department of Applied
Mathematics, University of Calcutta, for his continuous help,
suggestions and guidance during preparation of this manuscript. I
like to thank {\bf CSIR}, India for funding this research work. My
special thank goes to {\bf ICTP}, Trieste, Italy for giving me an
opportunity for two months visit and providing some excellent
facilities and information, which has facilitated to carry out
this work.
\section{Appendix-A:} We have chosen a frame in which
$$\overrightarrow{q}=\overrightarrow{q}_{1}+\overrightarrow{q}_{2}\eqno{(A1)}$$
In this frame we obtain
$$\frac{q^{2}}{2}+(p'q)=\frac{(p^{'0}-p^{0})(p^{'0}+p^{0})}{2}\eqno{(A2)}$$
$$\frac{q^{2}}{2}-(pq)=-\frac{(p^{'0}-p^{0})(p^{'0}+p^{0})}{2}\eqno{(A3)}$$
Thus,
$$\sum\mid\frac{(p'J)}{q^{2}/2+(p'q)}+\frac{(pJ)}{q^{2}/2-(pq)}\mid^{2}
=\frac{4}{(p^{'0}-p^{0})^{2}(p^{'0}+p^{0})^{2}}\sum\mid(p'-p)J\mid^{2}\eqno{(A4)}$$
Let us assume
$$P=p-p'=q-k\eqno{(A5)}$$
we have,
$$\sum\mid(PJ)\mid^{2}=\sum\mid\overline{u}_{\nu}(q_{1})P^{\mu}\gamma_{\mu}(1-\gamma_{5})
v(q_{2})\mid^{2}$$
$$\hspace{1.5cm}=\frac{2}{m_{\nu}^{2}}[2(q_{1}P)(q_{2}P)-(q_{1}q_{2})P^{2}]$$
$$\hspace{6cm}=\frac{2}{m_{\nu}^{2}}[(m_{\nu}P^{0})^{2}+\{(1-2cos^{2}\alpha)
\mid\overrightarrow{q}_{1}\mid^{2}+(q_{1}^{0})^{2}\}\mid
\overrightarrow{P}\mid^{2}]\eqno{(A6)}$$ Now using
$$\int d^{3}q_{2}=\frac{4\pi}{3}\mid\overrightarrow{q}_{2}\mid^{3}
=\frac{4\pi}{3}\mid\overrightarrow{q}_{1}\mid^{3}\eqno{(A7)}$$ and
$$d^{3}q_{1}=\mid\overrightarrow{q}_{1}\mid^{2}
d\mid\overrightarrow{q}_{1}\mid sin\alpha d\alpha
d\phi\eqno{(A8)}$$ we can obtain
$$\int\sum\mid (PJ)\mid^{2}\frac{d^{3}q_{1}}{2q_{1}^{0}}\frac{d^{3}q_{2}}{2q_{2}^{0}}
\delta(2q_{1}^{0}-q^{0})$$
$$=\frac{2\pi^{2}}{3m_{\nu}^{2}}\int\int_{\alpha=0}^{\pi}
\frac{\mid\overrightarrow{q}_{2}\mid^{4}}{q_{1}^{0}}[(m_{\nu}P^{0})^{2}+\{(1-2cos^{2}\alpha)
\mid\overrightarrow{q}_{1}\mid^{2}+(q_{1}^{0})^{2}\}\mid
\overrightarrow{P}\mid^{2}]\delta(q_{1}^{0}-\frac{q_{0}}{2})dq_{1}^{0}sin\alpha
d\alpha d\phi$$
$$=\frac{4\pi^{2}}{3m_{\nu}^{2}}\int\frac{\mid\overrightarrow{q}_{2}\mid^{4}}{q_{1}^{0}}
[(m_{\nu}P^{0})^{2}+\{\frac{\mid\overrightarrow{q}_{1}\mid^{2}}{3}+(q_{1}^{0})^{2}\}\mid
\overrightarrow{P}\mid^{2}]\delta(q_{1}^{0}-\frac{q_{0}}{2})dq_{1}^{0}$$
$$\approx \frac{\pi^{2}}{18m_{\nu}^{2}}(p^{0}-p^{'0})^{5}\mid
\overrightarrow{p}-\overrightarrow{p}'\mid^{2}\eqno{(A9)}$$ We
have assumed $m_{\nu}\ll q_{1}^{0}<q_{0}$ and used the following
criteria
$$m_{\nu}\longrightarrow 0$$
$$P^{0}=q^{0}=p^{0}-p^{'0}$$
$$\overrightarrow{P}=\overrightarrow{k}=\overrightarrow{p}-\overrightarrow{p}'$$
since $k$ is space like, whereas $q$ is time like in our chosen
frame.\\\indent Now introducing normalized factors and also using
$(A4)$ we obtain
$$\int\sum\mid\frac{(p'J)}{q^{2}/2+(p'q)}+\frac{(pJ)}{q^{2}/2-(pq)}\mid^{2}
\frac{d^{3}q_{1}d^{3}q_{2}}{(2\pi)^{3}2q^{0}_{1}(2\pi)^{3}2q^{0}_{2}}(2\pi)
\delta(q^{0}-q^{0}_{1}-q^{0}_{2})$$
$$\approx\frac{1}{18(2\pi)^{3}m_{\nu}^{2}}(p^{0}-p^{'0})^{3}[\frac{\mid\overrightarrow{p}
-\overrightarrow{p}'\mid}{p^{0}+p^{'0}}]^{2}\eqno{(A10)}$$ This is
same as the equation (2.11).
\section{Appendix-B} In presence of magnetic field the phase space
factor is replaced by the following relation \cite{Roulet}
$$\frac{2}{(2\pi)^{3}}\int d^{3}p=\frac{1}{(2\pi)^{2}}\sum_{n=0}^{n_{max}}g_{n}
\int dp_{z}\eqno{(B1)}$$ where $g_{n}$ represents degeneracy
factor of the Landau levels i.e.
$$g_{0}=1,\hspace{4cm} g_{n}=2\hspace{2cm}(n\geq 1)\eqno{(B2)}$$
The maximum Landau level $n_{max}$ can be obtained from the
following relation
$$n_{max}=\frac{1}{2m_{e}^{2}}(\frac{H}{H_{c}})[(p^{0}_{n_{max}})^{2}-(p^{0})^{2}]\eqno{(B3)}$$
where,
$$(p^{0})^{2}=p_{z}^{2}+m_{e}^{2}\eqno{(B4)}$$
For $n_{max}<1$ we have,
$$H > \frac{1}{2m_{e}^{2}}[(p^{0}_{n_{max}})^{2}-(p^{0})^{2}]H_{c}\eqno{(B5)}$$
It shows that for a very high magnetic field only $n=0$ Landau
level would contribute in the phase space. In this article we
consider the environment is highly magnetized, which gives
$$(p^{0}_{n_{max}})^{2}-(p^{0})^{2}> 2m_{e}^{2}$$
and therefore
$$H > H_{c}$$
In that case the phase space factor will take the form
$$\int d^{3}p=\pi \frac{H}{H_{c}}m_{e}^{2}\int dp_{z}\eqno{(B6)}$$
It is same as the equation (2.12).\\\indent If the magnetic field
is comparatively lower the higher Landau levels contribute in the
phase space as per the condition (B3).

\pagebreak\noindent{\large\bf \quad Figure Caption :}
\vspace{0.2cm}\\ {\bf Figure-1:} Feynman diagram for the neutrino
bremsstrahlung process
in presence of magnetic field with Z boson exchange.\\
{\bf Figure-2:} Feynman diagram for the neutrino bremsstrahlung
process in presence of magnetic field with W boson exchange.
\begin{table}
\begin{tabular}{|c|ccc|c|}\hline
$T_{10}$&&&$log(\frac{A}{Z^{2}}\mathcal{E_{\nu}})$\\\hline
&&&Presence of magnetic field& Absence of magnetic field\\\hline &
$10^{14}$ & $10^{15}$ & $10^{16}$\\\hline
$0.8$ & $7\cdot51$ &$9\cdot51$&${\bf 11\cdot51}$ &$11\cdot39$ \\
$0.9$ & $7\cdot62$ &$9\cdot61$&$11\cdot62$ &$11\cdot69$ \\
$1$ & $7\cdot71$ &$9\cdot71$&$11\cdot71$ &$11\cdot97$ \\
$2$ & $8\cdot31$ & $10\cdot31$ & $12\cdot31$ & $13\cdot77$\\
$3$ & $8\cdot66$ & $10\cdot67$ & $12\cdot67$ & $14\cdot83$ \\
$4$ & $8\cdot91$ & $10\cdot91$ & $ 12\cdot91$ & $15\cdot58$ \\
$5$ & $9\cdot10$ & $11\cdot10$ & $13\cdot10$ & $16\cdot16$ \\
$6$ & $9\cdot26$ & $11\cdot26$ & $13\cdot26$ & $16\cdot64$ \\
$7$ & $9\cdot40$ & $11\cdot40$ & $ 13\cdot40$ & $17\cdot04$ \\
$8$ & $9\cdot51$ & $11\cdot51$ & $13\cdot51$ & $17\cdot39$\\
$9$ & $9\cdot62$ & $11\cdot62$ & $13\cdot62$ & $17\cdot69$ \\
$10$ & $9\cdot71$ & $11\cdot71$ & $13\cdot71$ &$17\cdot97$\\\hline
\end{tabular}
\caption{Logarithmic expression for energy loss rate at $\rho=
10^{15}$ $gm/cm^{3}$, and magnetic field $H=10^{16}$, $10^{15}$,
$10^{14}$ G due to the neutrino bremsstrahlung process in presence
and absence of magnetic field respectively in the temperature
range $0.8\times 10^{10} - 10^{11}$ K. The bold number indicates
that the former process dominates over the later.}
\end{table}
\end{document}